\documentclass[final,3p,times,twocolumn]{elsarticle}
\usepackage{amssymb}
\usepackage[fleqn]{amsmath}
\usepackage{amsthm}
\usepackage{subfigure}
\usepackage{graphicx}

\numberwithin{equation}{section}
\biboptions{numbers,sort&compress}
\usepackage{hyperref}

\journal{Physics Letters A}

\begin{document}

\begin{frontmatter}

\title{A systemic method to construct the high order nonlocal symmetries}

\author[label]{Xiangpeng Xin}
\author{Yong Chen\corref{cor1}\fnref{label}}

\address[label]{ Shanghai Key Laboratory of Trustworthy Computing, East China Normal University, Shanghai, 200062, china}
\ead{ychen@sei.ecnu.edu.cn}
\cortext[cor1]{Corresponding author.}

\begin{abstract}
We propose a systemic method of applying the auxiliary systems of original equations to find the high order nonlocal symmetries of nonlinear evolution equation. In order to validate the effectiveness of the method, some examples are presented.
\end{abstract}


\end{frontmatter}

\section{Introduction}

Since the Lie group theory was introduced by Sophus Lie\cite{Lie1}, the study of Lie group has always been an important subject in mathematics and physics. Using both classical and non-classical Lie group approaches\cite{Ovsiannikov1,Ibragimov1,Bluman1,Nucci1,Gandarias1,Dong1}, one can reduce dimensions of partial differential equations(PDEs) and construct the analytical solutions of these PDEs. Apart from the symmetries mentioned above which can be called local symmetry, there exist so-called nonlocal symmetries which entered the literature in the 80s of the last century largely through the work of P.J. Olver\cite{Olver1}. It is a kind of specific symmetry which infinitesimal coefficients contain nonlocal variables, such as potential variables, eigenfunction in Lax pairs, etc. To search for nonlocal symmetries of the nonlinear systems is an interesting work, because the nonlocal symmetries\cite{Bluman2,Bluman3,Bluman4,Galas1} enlarge the class of symmetries and they are connected with integrable models. Recently,  Lou, Hu and Chen's work\cite{Lou1,Hu1} should be exemplifying this so fully: Starting from nonlocal symmetries related to B\"acklund transformation (BT) of potential KdV and the nonlocal symmetries which related to the Darboux transformation for the well-known Korteweg-de Vries (KdV) equation, the nonlocal symmetry is localized by introducing suitable and simple auxiliary dependent variables to generate new solutions from old ones and to consider some novel group invariant solutions; the explicit analytic interaction solutions between cnoidal waves and solitary wave are fund, some other integrable models both in finite dimensions and infinite dimensions are generated under new nonlocal symmetry.

However, it can be difficult how to take effective method to find the nonlocal symmetries of nonlinear PDEs, let alone systemic and uniform method. In a number of cases the nonlocal symmetries may be obtained with the help of a recurrence operator\cite{Akhatov1}. But sometimes the recursion operators of given system are difficult to obtain. Bluman introduced the concept of potential symmetry\cite{Bluman2} PDE system by writing a given PDE(s) in a conserved form, a related system with potentials as additional dependent variables is obtained. F Galas\cite{Galas1} obtained the nonlocal Lie-B\"{a}cklund symmetries by introducing the pseudo-potentials as a auxiliary system.

A basic problem for the construction of nonlocal symmetries is the proper choice of nonlocal variables. They are defined by integrable systems of differential equations which relate the nonlocal variables to the original differential variable. The choice of these differential equations is made on the basis of some additional considerations. Above methods of selecting auxiliary systems are different, however, the methods of seeking nonlocal symmetries have common points, i.e. seeking the Lie point symmetries or the Lie-B\"{a}cklund symmetries of the original equations with the auxiliary systems. But, these methods may lose some important results, for example, the integral terms or the high order derivative terms of nonlocal variables in the infinitesimal coefficients of the nonlocal symmetries. So we want to give a systemic and uniform method to construct nonlocal symmetries by assuming the symmetry structure having the high order derivative terms of nonlocal variables.

This paper is arranged as follows: In Sect.2, we present a systemic method and give the detailed process of seeking the nonlocal symmetries. Some examples will be given in Sect.3. Finally, some conclusions and discussions are given in Sect.4.

\section{The systemic method of nonlocal symmetries}

We consider a system $\mathcal{F}$ of $n$-th order differential equations in $p$ independent and $q$ dependent variables is given as a system of equations,
\begin{equation}\label{hs-1}
\begin{array}{*{20}c}
   {\Delta _v (x,u^{(n)} ) = 0,} & {v = 1,2, \cdots ,l},  \\
\end{array}
\end{equation}
involving $x = (x^1 ,x^2 , \cdots ,x^p ),u = (u^1 ,u^2 , \cdots ,u^q )$, and the derivatives of $u$ with respect to $x$ up to order $n$. the function $\Delta _v (x,u^{(n)} ) = (\Delta _1 (x,u^{(n)} ), \cdots ,\Delta _l (x,u^{(n)} ))$ will be assumed to be smooth in their arguments. Let $X=\mathbb{R}^p$ be the space representing the independent variables.

Let
\begin{equation}\label{hs-2}
V = \xi ^p (x,u)\frac{\partial }{{\partial x^p }} + \eta ^q (x,u)\frac{\partial }{{\partial u^q }},
\end{equation}
be the infinitesimal generator of the Lie group of point transformations $\tilde x = F(x,u,\varepsilon ),\tilde u = G(x,u,\varepsilon )$.

Next, we describe the method of constructing the nonlocal symmetries as follows. For simplicity, we consider the case $p=2,q=1$, i.e. $(x^1,x^2)=(t,x)$.

\textbf{Step 1.} Choose the proper auxiliary systems. Usually, one can use the Lax pair, potential system, pseudo-potential, etc. with the following forms,
\begin{equation}\label{hs-3}
\begin{array}{*{20}c}
   \begin{array}{l}
 F_\alpha  (x,t,u,u_x ,u_t , \cdots ,\psi _x ,\psi _t , \\
 \psi _{xx} ,\psi _{xt} ,\psi _{tt} , \cdots ,\psi _{\lambda x} ,\psi _{\mu t} ) = 0, \\
 \end{array} & {\alpha  \in \mathbb{Z^+} }  \\
\end{array}
\end{equation}
where $\psi  = (\psi ^1 ,\psi ^2 , \cdots ,\psi ^\beta  )$ denote $\beta$ auxiliary variables and  $\psi _{\lambda x}$ denote $\lambda$th-order partial derivatives with respect to $x$, $\psi _{\mu t}$ denote $\mu$th-order partial derivatives with respect to $t$.

Let $U\simeq\mathbb{R} $ be the space representing the single coordinate $u$, the space $U_1$ is isomorphic to $\mathbb{R}^2$ with coordinates $(u_x,u_t)$. Similarly,$U_2 \simeq \mathbb{R}^3$ has the coordinates representing the second order partial derivatives of $u$, and in general, $U_k \simeq \mathbb{R}^{k+1}$, since there are $k+1$ distinct $k$-th order partial derivatives of $u$. Finally, the space $U^{(k)}=U\times U_1\times\cdots\times U_k$ with coordinates $U^{(k)}=(u;u_x,u_t;u_{xx},u_{xt},u_{tt};\cdots)$.

\textbf{Step 2.} In this step, we prolong the basic space $X\times U$ to the space $X\times U^{(n)}$ , with coordinates$(x,t,u ,u , u _x ,u _t , \cdots) $. The $n$-th prolongation of $V$, denoted $\tilde V^{(n)}$, will be a vector field on the $n$-jet space $X\times U^{(n)}$. The vector field in general take the form
\begin{equation}\label{hs-4}
\tilde V^{(n)}  = \sum\limits_{i = 1}^2 {\xi ^i \frac{\partial }{{\partial x^i }}}  + \sum\limits_L {\eta ^L \frac{\partial }{{\partial u_L }}}.
\end{equation}

Here, we give a different definition of coefficients, i.e. the coefficient $\xi^i,\eta^L$ all depend on the variables $(x,t,u, \cdots ,\psi ,\psi,\psi _x ,\psi _t, \cdots , \psi _{\lambda x}, \psi _{\mu t} )$.  $\eta ^0=\eta$ and $\eta ^L $ have the form,
\begin{equation}\label{hs-5}
 \eta ^L  = D_L u - \sum\limits_{i = 1}^2 {u_L D_L \xi ^i }.
\end{equation}

\textbf{Remark 1:} The prolongation of vector fields show that this kind of symmetries neither classical Lie point symmetries nor Lie-B\"{a}cklund symmetries because it dependent on the auxiliary variables and their high order partial derivatives. More results may be obtained if we assume the coefficients $\xi^i,\eta^L$ have integral terms of the auxiliary variable i.e., they are the functions of $( x,t,u,\cdots, \int {\psi dx}, \cdots ) $.

\textbf{Step 3.} In order to seek the nonlocal symmetries, we should solve the following equations,
\begin{equation}\label{hs-6}
\tilde V^{(n)} \Delta _v (x,u^{(n)} )\left| {_{\scriptstyle \Delta _v (x,u^{(n)} ) = 0 \hfill \atop
  \scriptstyle Eqs.(3) \hfill} } \right. = 0.
\end{equation}

Using above equation, one can obtain a large number of elementary determining equation for the coefficient functions. Those determining equations can be solved and the general solution will determine the most general symmetry of the system.

\section{Examples}
\textbf{Example 1:} The well known Boussinesq equation\cite{Bogdanov1,Clarkson1},

\begin{equation}\label{hs-7}
u_{tt}  + (u^2 )_{xx}  + \frac{1}{3}u_{xxxx}  = 0,
\end{equation}
corresponding Lax pair of Eq.(\ref{hs-7}) has the form,
\begin{equation}\label{hs-8}
\psi _{xxx}= - \frac{3}{2}u\psi _x  - (\frac{3}{4}u_x +\frac{3}{4}\partial _x^{ - 1} u_t )\psi,
\end{equation}
\begin{equation}\label{hs-9}
\psi _t  =  - \psi _{xx} - u\psi,
\end{equation}
and its adjoint version is
\begin{equation}\label{hs-10}
\phi _{xxx}=-\frac{3}{2}u\phi _x  - (\frac{3}{4}u_x  - \frac{3}{4}\partial _x^{ - 1} u_t )\phi,
\end{equation}
\begin{equation}\label{hs-11}
\phi _t  = \phi _{xx}  + u\phi.
\end{equation}

That is to say the integrable conditions of Eqs.(\ref{hs-8}),(\ref{hs-9}) and (\ref{hs-10}),(\ref{hs-11}), $\psi _{xxxt}  = \psi _{txxx} $ and $\phi _{xxxt}  = \phi _{txxx} $ are just the Boussinesq equation (\ref{hs-7}).

Applying the Lax pair and its adjoint Lax pair of the Boussinesq equation as the auxiliary systems. Then, the vector field take the form,

\begin{equation}\label{hs-11-1}
V = \xi ^1 \frac{\partial }{{\partial x}} + \xi ^2 \frac{\partial }{{\partial t}} + \eta \frac{\partial }{{\partial u}}.
\end{equation}

One can prolong the basic space $V$ to the space $X\times U^{(4)}$  and obtain the prolongation of $V$,

\begin{equation}\label{hs-12}
\begin{array}{l}
 \tilde V  = \xi ^1 \frac{\partial }{{\partial x}} + \xi ^2 \frac{\partial }{{\partial t}} + \eta  \frac{\partial }{{\partial u}} + \eta ^{x} \frac{\partial }{{\partial u_x }} \\
 {\rm{  \quad       }}+ \eta ^{xx} \frac{\partial }{{\partial u_{xx} }}  + \eta ^{tt} \frac{\partial }{{\partial u_{tt} }} + \eta ^{xxxx} \frac{\partial }{{\partial u_{xxxx} }}
 \end{array}
\end{equation}
where the coefficients of $\tilde V$ all depend on the variables $(x,t,u,u_x,u_{xx},\psi ,\phi ,\psi _x ,\phi _x,\psi _{xx} ,\phi _{xx}  )$, and
\begin{equation}\label{hs-13}
\begin{array}{l}
 \eta ^t  = D_t (\eta  - \xi ^1 u_x  - \xi ^2 u_t ) + \xi ^1 u_{xt}  + \xi ^2 u_{tt} , \\
 \eta ^x  = D_x (\eta  - \xi ^1 u_x  - \xi ^2 u_t ) + \xi ^1 u_{xx}  + \xi ^2 u_{tx} , \\
 {\rm{  \qquad \qquad \qquad \qquad }} \vdots  \\
 \eta ^{xxxx}  = D_{xxxx} (\eta  - \xi ^1 u_x  - \xi ^2 u_t ) + \xi ^1 u_{xxxxx}  + \xi ^2 u_{txxxx} , \\
 \end{array}
\end{equation}

 Applying $\tilde V$ to Eq.(\ref{hs-7}), one can obtain the infinitesimal criterion (\ref{hs-6}) to be,
\begin{equation}\label{hs-14}
\eta ^{tt}  + 4u_x \eta ^{x}  + 2\eta u_{xx}  + 2u\eta ^{xx}  + \frac{1}{3}\eta ^{xxxx}  = 0.
\end{equation}

Substituting the general formulaes (\ref{hs-13}) in to (\ref{hs-14}), replacing $u_{tt}, \psi_{xxx}, \psi_{t}, \phi_{xxx}, \phi_{t}$ by Eqs.(\ref{hs-7}),(\ref{hs-8}),(\ref{hs-9}),(\ref{hs-10}) and (\ref{hs-11}). We get the determining equations for the functions $\xi^1,\xi^2,\eta$. Calculate by computer algebra, the general solutions of them take the form,
\begin{equation}\label{hs-15}
V = (\frac{1}{2}c_1 x + c_3 )\frac{\partial }{{\partial x}} + (c_1 t + c_2 )\frac{\partial }{{\partial t}} + (c_1 u - c_4 (\psi \varphi )_x )\frac{\partial }{{\partial u}}.
\end{equation}

\textbf{Remark 2:} The vector field (\ref{hs-15}) contains two parts, $V^1 = (\frac{1}{2}c_1 x + c_3 )\frac{\partial }{{\partial x}} + (c_1 t + c_2 )\frac{\partial }{{\partial t}} +  c_1 u\frac{\partial }{{\partial u}}$ and $V^2=-c_4 (\psi \varphi )_x \frac{\partial }{{\partial u}} $. One can see that the first part is the classical Lie point symmetry, and the second part is nonlocal symmetry. Therefore, both the general local symmetries and nonlocal symmetries can be obtained by this method. The second part is the same result in Ref.\cite{Hu2} which obtained using the Darboux transformation.

\textbf{Example 2:} The coupled KdV system\cite{Hu3,Fan1},
\begin{equation}\label{hs-16}
\begin{array}{l}
 u_t  =  - 6vv_x  + 6uu_x  - u_{xxx} , \\
 v_t  = 6uv_x  + 6vu_x  - v_{xxx},  \\
 \end{array}
\end{equation}
the Lax pair for Eq.(\ref{hs-16}) is as follows,
\begin{center}
$\begin{array}{l}
 \phi _{1xx}  = v\phi _2  + u\phi _1  - \lambda \phi _1 , \\
 \phi _{2xx}  = u\phi _2  - v\phi _1  - \lambda \phi _2 , \\
 \phi _{1t}  =  - 4\phi _{1xxx}  + 6v\phi _{2x}  + 6u\phi _{1x}  + 3v_x \phi _2  + 3u_x \phi _1 , \\
 \phi _{2t}  =  - 4\phi _{2xxx}  + 6u\phi _{2x}  - 6v\phi _{1x}  + 3u_x \phi _2  - 3v_x \phi _1 . \\
 \end{array}$
\end{center}

The vector field take the form
\begin{equation}\label{hs-17}
V = \xi ^1 \frac{\partial }{{\partial x}} + \xi ^2 \frac{\partial }{{\partial t}} + \eta ^1 \frac{\partial }{{\partial u}} + \eta ^2 \frac{\partial }{{\partial v }} .
\end{equation}

Using the formulas (\ref{hs-4}), one can prolong the space $V$ to the space  $X \times U^{(3)} \times V^{(3)}$, here we omit. Applying the prolonged vector field and following the step 3, one can get the the general solutions,

\begin{equation}\label{hs-18}
\begin{array}{l}
 V = (\frac{{c_1 x}}{3} + c_3 t + c_4 )\frac{\partial }{{\partial x}} + (c_1 t + c_2 )\frac{\partial }{{\partial t}} \\
 {\rm{ \quad   }} + [\frac{{c_6 }}{2}(\phi _1^2  - \phi _2^2 )_x  - c_5 (\phi _1 \phi _2 )_x  + \frac{{2c_1 u}}{3} + \frac{{c_3 }}{6}]\frac{\partial }{{\partial u}} \\
 {\rm{ \quad   }} - [c_6 (\phi _1 \phi _2 )_x  - \frac{{c_5 }}{2}(\phi _2^2  - \phi _1^2 )_x  - \frac{{2c_1 v}}{3}]\frac{\partial }{{\partial v}}. \\
 \end{array}
\end{equation}

From above vector field $V$, one can see that this is composed of classical symmetries and nonlocal symmetries.

\textbf{Example 3:} The KP equation\cite{Lou2,Weiss1} has the following form,
\begin{equation}\label{hs-19}
u_{xt}  - 6u_x^2  - 6uu_{xx}  + u_{xxxx}  + 3u_{yy}  = 0.
\end{equation}

It is well known that the KP equation possesses the Lax pair and the adjoint Lax pair,
\begin{equation}\label{hs-20}
\begin{array}{l}
 \psi _{xx}  = u\psi  - \psi _y , \\
 \psi _t  =  - 4\psi _{xxx}  + 6u\psi _x  + 3(u_x  - \int {u_y dx} )\psi,  \\
 \phi _{xx}  = u\phi  + \phi _y , \\
 \phi _t  =  - 4\phi _{xxx}  + 6u\phi _x  + 3(u_x  + \int {u_y dx} )\phi.  \\
 \end{array}
\end{equation}

Let the vector field of Eq.(\ref{hs-19}) takes the form,

\begin{equation}\label{hs-20-1}
V = \xi ^1 \frac{\partial }{{\partial x}} + \xi ^2 \frac{\partial }{{\partial y}}+ \xi ^3 \frac{\partial }{{\partial t}} + \eta \frac{\partial }{{\partial u}}  .
\end{equation}

Using the prolonged vector field and following the step 3, one can get the solution,
\begin{equation}\label{hs-21}
\begin{array}{l}
 V = (\frac{x}{3}F_{1t}  - \frac{{y^2 }}{{18}}F_{1tt}  - \frac{y}{6}F_{2t}  - 6F_3  + c_2 )\frac{\partial }{{\partial x}} \\
 {\rm{ \quad   }} + (\frac{{2y}}{3}F_{1t}  + F_2 )\frac{\partial }{{\partial y}} + F_1 \frac{\partial }{{\partial t}} + (\frac{{2u}}{3}F_{1t}  + \frac{x}{{18}}F_{1tt}  \\
 {\rm{ \quad   }} - c_1 (\psi \phi )_x  - \frac{{y^2 }}{{108}}F_{1ttt}  - \frac{y}{{36}}F_{2t}  - F_{3t} )\frac{\partial }{{\partial u}} \\
 \end{array}
\end{equation}

Because the coefficient of $\frac{\partial }{{\partial u}}$ contains the nonlocal variables $\psi ,\phi$, so it is a nonlocal symmetry of KP equation.

\textbf{Example 4:} The AKNS equation\cite{Hu2,Zeng1}
\begin{equation}\label{hs-22}
\begin{array}{l}
 u_t  =  - iu_{xx}  + 2iu^2 v, \\
 v_t  = iv_{xx}  - 2iv^2 u. \\
 \end{array}
\end{equation}

the Lax pair for Eq.(\ref{hs-22}) is as follows,
\begin{equation}\label{hs-23}
\begin{array}{l}
 \left( {\begin{array}{*{20}c}
   {\phi _{1x} }  \\
   {\phi _{2x} }  \\
\end{array}} \right) = \left( {\begin{array}{*{20}c}
   { - i\lambda } & u  \\
   v & {i\lambda }  \\
\end{array}} \right)\left( {\begin{array}{*{20}c}
   {\phi _1 }  \\
   {\phi _2 }  \\
\end{array}} \right) \\
 \left( {\begin{array}{*{20}c}
   {\phi _{1t} }  \\
   {\phi _{2t} }  \\
\end{array}} \right) = \left( {\begin{array}{*{20}c}
   {2i\lambda ^2  + iuv} & { - 2\lambda u - iu_x }  \\
   { - 2\lambda v + iv_x } & { - 2i\lambda ^2  - iuv}  \\
\end{array}} \right)\left( {\begin{array}{*{20}c}
   {\phi _1 }  \\
   {\phi _2 }  \\
\end{array}} \right) \\
 \end{array}
\end{equation}

Using the same method we can obtain the nonlocal symmetries take the form,
\begin{equation}\label{hs-24}
\begin{array}{l}
 V = (c_1 t + c_3 x + c_2 )\frac{\partial }{{\partial x}} + (2c_3 t + c_4 )\frac{\partial }{{\partial t}} \\
 {\rm{  \quad   }} - (\frac{{4c_3  + 2c_6  + c_1 ix}}{2}u - c_5 \phi _1^2 )\frac{\partial }{{\partial u}} \\
 {\rm{   \quad    }} + (\frac{{2c_6  + c_1 ix}}{2}v + c_5 \phi _2^2 )\frac{\partial }{{\partial v}} \\
 \end{array}
\end{equation}

\textbf{Remark 3:} From the above examples, we can see that this method is very effective for searching the nonlocal symmetries of DE(s). The method is also suitable for other models which have Lax pairs, pseudo-potentials etc.

\section{Conclusions}
To search for nonlocal symmetries of integrable DEs and apply the nonlocal symmetries to construct explicit solutions are both of considerable interest and value. There are many kinds of methods to search for the nonlocal symmetry, and the basic problem of those methods is the proper choice of nonlocal variables. Here a systemic method of applying the auxiliary systems of original equations to find the high order nonlocal symmetry of nonlinear evolution equation was presented. Through some examples, we can learn that this method can get nonlocal symmetries effectively.

Moreover, how to use the nonlocal symmetry to build similarity solutions is another important problem. In the Ref.\cite{Hu1} the nonlocal symmetry can be localized when the three potentials,the spectral function $\psi$, the $x$ derivatives of the spectral functions $\psi_1=\psi_x$ and the singularity manifold function $p=\int {\psi^2 dx}$, are introduced. In this case, the primary nonlocal symmetry is equivalent to a Lie point symmetry of a prolonged system, on the basis of which one can find nonlocal groups as well as the explicit similarity solutions.

In this paper, we give some high order nonlocal symmetries which can be localized using above method. The applications of seeking exact solution of these nonlinear systems are worthy of further study. These conclusions may be useful for the explanation of some practical physical problems.

\section*{Acknowledgments:}
We would like to thank Prof. S.Y. Lou for his enthusiastic guidance and helpful discussions. This work is supported by the National Natural Science Foundation of China (Nos. 11275072 and 11075055), Research Fund for the Doctoral Program of Higher Education of China (No. 20120076110024), Innovative Research Team Program of the National Natural Science Foundation of China (No. 61021004), Shanghai Leading Academic Discipline Project (No. B412) and National High Technology Research and Development Program (No. 2011AA010101).

%\bibliography{yqyang}



\end{document}